\documentclass[aps,showpacs,amssymb,amsmath,nofootinbib]{revtex4}

\usepackage{amsmath}

\usepackage{amssymb}

\def\beq{\begin{equation}}
\def\eeq{\end{equation}}

\def\bea{\arraycolsep .1em \begin{eqnarray}}
\def\eea{\end{eqnarray}}

\begin{document}

\title{Dynamically Generated Anomalous Magnetic Moment in Massless QED}
\author{Efrain J. Ferrer and Vivian de la Incera}
\address{Department of Physics, University of Texas at El Paso,
El Paso, TX 79968, USA}

\begin{abstract}

In this paper we investigate the non-perturbative generation of an anomalous magnetic moment for massless fermions in the presence of an external magnetic field. In the context of massless QED in a magnetic field, we prove that the phenomenon of magnetic catalysis of chiral symmetry breaking, which has been associated in the literature with dynamical mass generation, is also responsible for the generation of a dynamical anomalous magnetic moment. As a consequence, the degenerate energy of electrons in Landau levels higher than zero exhibits Zeeman splitting. We explicitly report the splitting for the first Landau level and find the non-perturbative Lande $\textit{g}$-factor and Bohr magneton. We anticipate that a dynamically generated anomalous magnetic moment will be a universal feature of theories with magnetic catalysis. Our findings can be important for condensed planar systems as graphene, as well as for highly magnetized dense systems as those forming the core of compact stars.

\pacs{11.30.Rd, 13.40.Em, 21.65.Qr, 81.05.Uw}
\end{abstract}
\maketitle

\section{Introduction}

The quantum mechanical
description of charged fermions in a constant
magnetic field has attracted much
attention since the early developments of Quantum Electrodynamics (QED)
\cite{H-QED}-\cite{Johnson}. The partial Lorentz symmetry breaking produced by the
magnetic field is reflected in the quantum mechanical properties of
the charged fermions which behave as free particles along the
field, but have quantized momenta \cite{Landau} (characterized by a Landau level $l$) in the transverse direction. At the tree level, the Landau levels $l\geq1$ are degenerate with respect to the spin projections along and opposite to the magnetic field. However, this degeneracy is broken by radiative corrections \cite{Schwinger}, giving rise to a spin-field interaction term in the effective action and thus to an anomalous magnetic moment.

The theory of the electron magnetic moment has historically played
an important role in the development of QED. As it is known, the
electron intrinsic magnetic moment $\boldsymbol{\mu}$ is related
to the spin vector $\boldsymbol{s}$ by
$\boldsymbol{\mu}=g\mu_{B}\boldsymbol{s}$, where
$\mu_{B}=e\hbar / 2mc$ is the Bohr magneton, and $g$ is the Lande
$g$-factor. One of the greatest triumphs of the Dirac theory
was the prediction of the value $g=2$ for the Lande $g$-factor of the electron in the non-relativistic limit.
However, this prediction was later challenged by more refined experimental measurements showing a value slightly larger than 2. The solution of this puzzle came only after Schwinger calculated the first-order
radiative correction to $\boldsymbol{\mu}$ due to the
electron-photon interactions \cite{Schwinger}. Schwinger's results
led to an anomalous magnetic moment with a correction to the
$g$-factor of order $\frac{g-2}{2}=\frac{\alpha}{2\pi}$, $\alpha$
being the fine-structure constant. Higher-order
radiative corrections to $g$ have subsequently led to a series in powers of
$\alpha/\pi$ \cite{Kinoshita-Lindquist} that is in excellent
agreement with the experiment.

For massless electrons, however, no anomalous magnetic moment can be
found through Schwinger's perturbative approach. The
problem is that an anomalous magnetic moment would break the chiral symmetry
of massless QED, but this symmetry is protected against any perturbatively
generated breaking term. However, no protection exists against non-perturbative breaking of
chiral symmetry. A non-perturbative chiral symmetry breaking mechanism is known to
exist in theories of massless fermions in a constant and uniform magnetic field \cite{MC}.
It is called the magnetic catalysis of chiral symmetry
breaking ($MC$$\chi$$SB$). The $MC$$\chi$$SB$ gives rise to a chiral condensate that in turns can
produce a dynamical fermion mass. Until recently, all the studies of $MC$$\chi$$SB$ focused
on the generation of a dynamical mass \cite{MC}-\cite{MC-Gen}, but
ignored the possibility of a dynamically generated anomalous magnetic moment. In a
recent letter \cite{Zeeman}, we reconsidered the $MC$$\chi$$SB$ in massless
QED and showed that in addition to the dynamical mass, the magnetic catalysis simultaneously
produces a dynamically generated anomalous magnetic moment. Our results gave rise to a non-perturbative Bohr
magneton proportional to the inverse of the dynamical mass, in analogy to the way in which the bare Bohr magneton
depends on the bare mass. In the present paper we develop in detail the calculations
that led to the magnetically catalyzed anomalous magnetic moment found in Ref. \cite{Zeeman}. We also discuss some elucidating points about the infrared dynamics of fermions in the lowest Landau level (LLL) and their role in the generation of the anomalous magnetic moment for fermions in higher Landau levels.

The plan of the paper is as follows. In Sec. 2, we present a brief historical account of the appearance of magnetic moment contributions in relativistic and non-relativistic theories of charged fermions, including a summary of our own findings for the case of massless fermions, which will be developed in detail in the rest of the paper. In Sec. 3, we calculate the fermion full propagator in massless QED in a magnetic field, considering the Dirac's structures for mass, magnetic moment and wave-function renormalization term. In Sec. 4, an infinite system of coupled Schwinger-Dyson (SD) equations for the fermion self-energy is derived within the ladder approximation. This system is then consistently solved in Sec. 5, where explicit solutions of the dynamical quantities are found in the first and in the LLL. The dispersion relations for fermions at different LLs in the chiral-condensate phase are calculated in Sec. 6. In particular, the Zeeman effect for fermions in the first LL is analyzed and the corresponding non-perturbative Lande's $\textit{g}$-factor and Bohr magneton are identified. In Sec. 7, we calculate the chiral condensate and find the connection between this order parameter and the rest energy of the electrons in the LLL, which is given by the sum of the dynamical mass and the magnetic energy due to the anomalous magnetic moment term. This result reflects the infrared origin of the condensation phenomenon and underlines the fact that both parameters, the mass and the anomalous magnetic moment, can be induced once the chiral symmetry is broken by the condensate. Possible applications of the outcomes of this paper for planar condensed matter systems as graphene and for dense astrophysical systems as compact stars or magnetars are outlined in Sec. 8. Finally, using a chiral-spin representation for fermions in the LLL, it is shown in the Appendix that the dynamics of such fermions reduces to that of free particles in a (1+1)-dimensional space with an induced rest-energy that depends non-perturbatively on the coupling constant and applied magnetic field.

\section{Magnetic Moment Historical Review}\label{zeeman-eff-gener}

For the sake of understanding we will briefly review in this section the role that
spin-field interactions have played in
non-relativistic and relativistic theories of spin-$\frac{1}{2}$ charged particles in a magnetic field. We discuss massive and massless QED and explain why in the massless theory we cannot follow the perturbative approach that gave rise to an anomalous magnetic moment in the massive situation. It will become clear throughout the paper that a non-perturbative method is required in the massless case because the anomalous magnetic moment there will be closely connected to the breaking of the original chiral symmetry, hence it has to be dynamically generated through some non-perturbative mechanism.

\subsection{Non-Relativistic Case}

Let us start with the case of a non-relativistic spinless charged particle in the presence
of a constant and uniform magnetic field $H$. Assuming a magnetic field pointing along the $x_3$-direction, the energy eigenvalues $E_{S}$ of the Schrodinger
Hamiltonian
\begin{equation}
H_{S}=\frac{1}{2m}(\mathbf{p}-e\mathbf{A})^2
 \label{2}
\end{equation}
are given by
\cite{Landau, Lifshitz}
\begin{equation}
E_{S}=\frac{p_{3}^2}{2m}+\frac{|eH|}{m}(n+\frac{1}{2}) \qquad
n=0,1,2,... , \label{3}
\end{equation}
Notice that the particle can freely move along the field direction, while in the plane perpendicular
to the field its motion is confined to quantum orbits labeled by the discrete numbers $n$. Without
loss of generality, we can assume from now on that $|eH|=eH$.

For charged fermions, the Schrodinger Hamiltonian needs to include a spin-field interaction
term originally introduced by Pauli as
\begin{equation}
H_{P}=\frac{1}{2m}(\mathbf{p}-e\mathbf{A})^2-\boldsymbol{\mu} \cdot \mathbf{H} , \label{4}
\end{equation}
where
\begin{equation}
\boldsymbol{\mu}=g\mu_{B}\mathbf{s}
\label{5}
\end{equation}
is the particle intrinsic magnetic moment. Here $\mu_{B}$ is the Bohr magneton, $\textbf{s}$ the spin
operator given in terms of the Pauli matrices as
$\mathbf{s}=\boldsymbol{\sigma}/2$, and $g$ the Lande $g$-factor. For bare electrons $g=2$.

The corresponding energy eigenvalues are
\begin{equation}
E_{P}=\frac{p_{3}^2}{2m}+\frac{eH}{m}(n+\frac{1}{2})-\mu_{B}\sigma H
\qquad n=0,1,2,... ,\qquad \sigma=\pm 1 \label{6}
\end{equation}
Here $\sigma$ denotes spin projections along $(+)$ and opposite $(-)$ to
the field direction. The last equation can be rewritten as
\begin{equation}
E_{P}=\frac{p_{3}^2}{2m}+2\mu_{B}(n+\frac{1}{2}-\frac{\sigma}{2})H=\frac{p_{3}^2}{2m}+2\mu_{B}lH
\label{7}
\end{equation}
where
\begin{equation}\label{13}
l=n+\frac{1}{2}-\frac{\sigma}{2}, \qquad l=0,1,2,...
\end{equation}
The number $l$ is known as the Landau level number. Notice the double degeneracy
of all the $l\neq 0$ due to the two spin projections contributing to each
of them. The LLL ($l=0$) is not degenerate because there
is only one combination of the non-negative integer $n$ and spin projection $\sigma$ that can produce $l=0$.
As we will see below, the degeneracy of $l\neq 0$ can be broken by radiative
corrections which give rise to an anomalous magnetic moment
term in the electron self-energy.

\subsection{Relativistic Case}

In the relativistic case we should start from the Dirac equation in the presence of an external
magnetic field
\begin{equation}
(\Pi_{\mu}\gamma^{\mu}-m)\psi=0 \label{8}
\end{equation}
where the field is introduced through the covariant
derivative
\begin{equation}
\Pi_{\mu}=i\partial_{\mu}-eA_{\mu} \label{9}
\end{equation}

Assuming again a uniform and constant magnetic field along the $x_3$ direction, and using the gauge
$A_2=Hx_1$, $A_0=A_1=A_3=0$, Eq. (\ref{8}) can be solved for $\psi$. It follows from (\ref{8}) that the energy eigenvalues for the relativistic spin-$\frac{1}{2}$
particle are found \cite{Johnson} to be
\begin{equation}
E_{R}=\pm\sqrt{m^2+p_{3}^2+2eH(n+\frac{1}{2}-\frac{\sigma}{2})}
\qquad n=0,1,2,... ,\qquad \sigma=\pm 1 \label{10}
\end{equation}
with ($+$) for particles and
 ($-$) for antiparticles. In terms of the Landau levels (\ref{13}) the relativistic energy becomes
\begin{equation}
E_{R}=\pm\sqrt{m^2+p_{3}^2+2eHl}
\qquad l=0,1,2,...  \label{10-2}
\end{equation}
Clearly the double
spin degeneracy of the $l\neq0$ states is also present for relativistic fermions.

The non-relativistic
energy can be easily recuperated by taking the limit $2eH/m^2\ll 1$ and
$p_3^2/m^2\ll 1$ in (\ref{10-2})  and subtracting the rest energy $m$,
\begin{equation}\label{11}
E_{NR}=\lim_{2eH/m^{2},p^{2}_{3}/m^{2}\rightarrow 0} E_{R}-m=
\lim_{2eH/m^{2},p^{2}_{3}/m^{2}\rightarrow 0} m\sqrt{1+\frac{p_{3}^2}{m^2}+\frac{2eH}{m^2}l}-m
\simeq
\frac{p_{3}^2}{2m}+2\mu_B lH .
\end{equation}

Since the spin is automatically incorporated in the relativistic treatment, the
non-relativistic limit leads directly to the Pauli theory (\ref{7}). Even more, it naturally gives
that a unit of spin angular momentum ($\sigma /2$) interacts with
the magnetic field with a coupling of $2\mu_{B}$, that is, the Dirac
theory automatically produces the correct $g$-factor of 2, something that was a puzzle at those times.

\subsection{Anomalous Magnetic Moment}

Despite Dirac theory's success in predicting $g=2$, this result was later challenged by
experimental findings of $g>2$ for electrons/positrons.  The solution to this puzzle was provided
by Schwinger. In a classical paper on the topic \cite{Schwinger}, Schwinger calculated the one-loop contribution to the fermion
self-energy in a weak magnetic field that led to an anomalous magnetic moment \begin{equation}
\boldsymbol{\mu}_{A}=(g'-2)\mu_{B}\boldsymbol{s}
\label{14} \end{equation}
Accordingly, the Lande $g$-factor was modified as \begin{equation}
g'=2(1+\frac{\alpha}{2\pi}) \label{15} \end{equation}
in good agreement with the experiment.

Schwinger's result contained the first order correction in $\alpha/\pi$.
Subsequent higher-order corrections to $g$ give rise to a series in powers of
$\alpha/\pi$ \cite{Kinoshita-Lindquist}. The corrections up to the eighth order
has shown an agreement with the experimental value that is good to one
part in $10^{12}$ \cite{Mohr}.

Taking into account Schwinger's anomalous magnetic
moment, the Dirac equation acquires an extra structure $\sigma_{\mu\nu}F^{\mu\nu}$ (with $\sigma_{\mu\nu}=\frac{i}{2}[\gamma_{\mu}\gamma_{\nu}]$), so
\begin{equation}
(\Pi_{\mu}\gamma^{\mu}-m+\kappa\mu_BH\Sigma_3)\psi=0 \label{16}
\end{equation}
where $\kappa$=$\alpha /2\pi$ and $\Sigma_3=i\gamma_1 \gamma_2$ is the spin operator. The corresponding
relativistic-particle energy  \cite{Yildiz} is
\begin{equation}
E_{l,\sigma}^2=[(m^2+2eHl)^{1/2}-\mu_B\kappa H\sigma]^2+p_{3}^2,
\quad l=0,1,2,... ,\quad \sigma=\pm 1 \label{17}
\end{equation}
Hence, once the anomalous magnetic moment contribution is considered
the spin degeneracy is removed since the energy (\ref{17})
explicitly depends on the spin projection $\sigma$. It is timely to clarify here a mistake appearing in Ref. \cite{Wang}, where it was argued that a magnetic moment term ($\sigma_{\mu\nu}F^{\mu\nu}$) could not be present in the self-energy because it would give rise to an energy that would depend on the orientation of the magnetic field. It is obvious from (\ref{17}), that no matter what the direction of the magnetic field is, the energy of the particle with spin oriented along or opposite to the field will not change. Notice that a $180^\circ$ rotation equally affects the magnetic field and the spin, leaving  the magnetic moment contribution to the energy $-\boldsymbol{\mu}\cdot \textbf{H}$ unchanged.

As was done in Eqs. (\ref{11}), we can take the non-relativistic limit of (\ref{17}) to find
\begin{eqnarray}\label{18}
E_{NR}=\frac{p_{3}^2}{2m}+2(n+\frac{1}{2})\mu_B H-2(1+\kappa)\mu_B
\frac{\sigma}{2}H=\qquad \qquad \qquad \qquad \nonumber
\\
=\frac{p_{3}^2}{2m}+2(n+\frac{1}{2})\mu_B H-g'\mu_B
\frac{\sigma}{2}H=\frac{p_{3}^2}{2m}+2l\mu_B H-\kappa \mu_B \sigma H, \end{eqnarray}
This result shows that the Lande $g$-factor depends on the fine-structure
constant as pointed out in (\ref{15}), and that the
anomalous magnetic moment breaks the spin degeneracy of all LLs
with $l\geq 1$, hence producing the following energy splitting of the levels
\begin{equation}
\Delta E_{l}=2\kappa \mu_B H=(g'-2)\mu_B H  \label{18-2}
\end{equation}
The case with strong fields, i.e. fields about $B_c \sim 10^{13} G$, was considered in Refs.\cite{Strong Field1}-\cite{Strong Field2}. In those works, the authors treated the field exactly, (i.e. without expanding in powers of $H$), in the fermion one-loop self energy but kept a perturbative treatment in the coupling constant. In that approximation the energy splitting due to the anomalous magnetic moment no longer changes linearly with $H$, and besides it depends on $l$ \cite{Strong Field2}.

\subsection{Induced Magnetic Moment for Massless Fermions}

As previously shown, the electron magnetic moment is related through the
Bohr magneton $\mu_B$ to its charge-to-mass ratio. Hence, the origin
of the extra contribution due to the anomalous magnetic moment
appearing in quantum field theory can be understood from the fact
that there the electron is continuously self-interacting through
radiative interactions. Thus, part of the electron energy, and
consequently of the electron mass, will be transferred to the created
photon cloud. Therefore, as a consequence of the decrease of the
electron's mass, the corresponding magnetic moment will be
strengthened.

On the other hand, as we have already stressed in the Introduction, if one is interested in
exploring the appearance of an anomalous magnetic moment in a theory of
massless fermions, a perturbative, "a-la-Schwinger" approach is not possible anymore. Since an anomalous magnetic
moment would break the chiral symmetry of the massless theory, it can only be generated via
non-perturbative effects.

Henceforth, we are going to explore the dynamical generation of an anomalous magnetic
moment in the context of massless 4-dimensional QED in the presence of a uniform and constant magnetic field.
As we are going to show in the next sections, the formation of a chiral condensate through the
$MC$$\chi$$SB$ mechanism is responsible not just for the dynamical generation of a mass, but also
for the appearance of a dynamical magnetic moment.

Physically it is easy to understand the origin of the new dynamical
quantity. The chiral condensate carries non-zero magnetic moment,
since the particles forming the condensate have opposite spins and
opposite charges. Therefore, chiral condensation will inexorably
provide the quasiparticles with both a dynamical mass and a
dynamical magnetic moment. Symmetry arguments can help us also to
better understand this phenomenon. A magnetic moment term does not
break any additional symmetry that has not already been broken by a
mass term. Hence, once $MC\chi SB$ occurs, there is no reason why
only one of these parameters should be different from zero.

We will show below that a very important consequence of the
dynamically generated magnetic moment is a splitting in the electron
energy spectrum that can be interpreted as a non-perturbative Zeeman
effect. In the LLL, since only electrons with one spin projection
are allowed, there is no energy degeneracy and therefore no
splitting can occur. However, for electrons in higher LLs the energy of
the degenerated spin states is splitted by the interaction of the
induced magnetic moment with the applied field. The corresponding
energy splitting can be conveniently written in the well known form
of the Zeeman energy splitting for the two spin projections as
\begin{equation}\label{Zeeman-splitting}
\Delta E=2\widetilde{\kappa}\widetilde{\mu}_B H
\end{equation}
where $\widetilde{\kappa}$ and $\widetilde{\mu}_B$ are the
non-perturbative Lande $g$-factor and Bohr magneton respectively. For
electrons in the first LL they are given by \cite{Zeeman}
\begin{equation}\label{Zeeman-splitting-const}
\widetilde{\kappa}=e^{-2\sqrt{\pi /\alpha}},\quad \widetilde{\mu}_B
=\frac{e}{2M^1}
\end{equation}
 Worth to highlight in the above results are the non-perturbative dependence on the coupling constant $\alpha$ in the Lande $g$-factor and the Bohr magneton's dependence on the dynamically induced electron mass $M^1$.

\section{Electron Full Propagator in Momentum Space
}\label{full-propagator}

In this Section we are going to obtain the fermion's full propagator in QED with massless bare
fermions in the presence of a constant and uniform magnetic field. The full propagator obeys the following
equation

\begin{equation}\label{fullpropagator-eq}
[\Pi_\mu \gamma^\mu - \Sigma(x,x')]G(x,x')=\delta^4(x-x')
\end{equation}
The structure of the the self-energy $\Sigma(x,x')$ \cite{WT-identity}
\begin{equation}
\Sigma (x,x')=(Z_{\|}\Pi_\mu^{\|}
\gamma^\mu_{\|}+Z_{\bot}\Pi_\mu^{\bot} \gamma^\mu_{\bot} +M+\frac{T}{2}
\widehat{F}^{\mu \nu}\sigma_{\mu \nu})\delta^4(x-x')\label{Mass
Operator}
\end{equation}
contains the wave function's renormalization coefficients $Z_{\|}$ and $Z_{\bot}$, as well as mass $M$ and anomalous magnetic moment $T$ terms, all of which have to be determined self-consistently
as the solutions of the SD equations of the theory.

In (\ref{Mass
Operator}) $\widehat{F}^{\mu \nu}=F^{\mu
\nu}/H$ denotes the normalized electromagnetic strength tensor, with
$H$ the field strength. The external magnetic field breaks the rotational
symmetry of the theory, hence separating between longitudinal $p_{\|}\cdot \gamma^{\|}=p^{\nu }\widehat{F}_{\nu \rho}^{\ast} \widehat{F}^{\ast \mu \rho}\gamma _{\mu }$ (for $\mu, \nu=0,3$),
and transverse $p_{\bot}\cdot \gamma^{\bot}=p_{\mu}\widehat{F}^{\mu\rho}\widehat{F}_{\rho\nu}\gamma^{\nu}$ (for $\mu, \nu=1,2$), modes. $\widehat{F}_{\mu\nu}^{ \ast} =\frac{1}{2H}\varepsilon _{\mu \nu \rho \lambda}F^{\rho \lambda }$ is
the dual of the normalized electromagnetic strength tensor
$\widehat{F}_{\mu\nu }$.

The transformation to  momentum space of (\ref{Mass Operator}) can
be done by using the so-called Ritus' method, originally developed
for fermions in \cite{Ritus:1978cj} and later extended to vector
fields in \cite{efi-ext}. In Ritus' approach, the transformation to
momentum space is carried out using the eigenfunctions
$E_{p}^{l}(x)$ of the asymptotic states of the charged fermions in a
uniform magnetic field
\begin{equation}\label{Ep}
 E_{p}^{l}(x)=E_{p}^{+}(x)\Delta(+)+E_{p}^{-}(x)\Delta(-)
\end{equation}
where
\begin{equation}
\Delta(\pm)=\frac{I\pm i\gamma^{1}\gamma^{2}}{2},
\label{Spin-projectors}
\end{equation}
are the spin up ($+$) and down ($-$) projectors, and
$E_{p}^{+/-}(x)$ are the corresponding eigenfunctions
\begin{eqnarray}\label{E-x}
E_{p}^{+}(x)=N_{l}e^{-i(p_{0}x^{0}+p_{2}x^{2}+p_{3}x^{3})}D_{l}(\rho),\qquad
\nonumber
\\
E_{p}^{-}(x)=N_{l-1}e^{-i(p_{0}x^{0}+p_{2}x^{2}+p_{3}x^{3})}D_{l-1}(\rho)
\end{eqnarray}
with normalization constant $N_{l}=(4\pi eH)^{1/4}/\sqrt{l!}$, and
$D_{l}(\rho)$ denoting the parabolic cylinder functions of argument
$\rho=\sqrt{2eH}(x_{1}-p_{2}/eH)$, and index given by the Landau
level numbers $l=0,1,2,...$.

The $E_p^l$ functions satisfy the orthogonality condition
\cite{Wang}
\begin{equation}
\int d^{4}x \overline{E}_{p}^{l}(x)E_{p'}^{l'}(x)=(2\pi)^4
\widehat{\delta}^{(4)}(p-p')\Pi(l) \ , \label{orthogonality}
\end{equation}
with $\overline{E}_{p}^{l}\equiv \gamma^{0}
(E_{p}^{l})^{\dag}\gamma^{0}$,
\begin{equation}
\widehat{\delta}^{(4)}(p-p')=\delta^{ll'} \delta(p_{0}-p'_{0})
\delta(p_{2}-p'_{2}) \delta(p_{3}-p'_{3}), \label{gen-delta}
\end{equation}
and
\begin{equation}
\Pi(l)=\Delta(+)\delta^{l0}+I(1-\delta^{l0}). \label{degeneracy}
\end{equation}

The spin structure of the $E_p$ functions (\ref{Ep}) is essential to satisfy
the eigenvalue equations
\begin{equation}
(\Pi\cdot\gamma)E^{l}_{p}(x)=E^{l}_{p}(x)(\gamma\cdot\overline{p}) \
, \label{eigenproblem}
\end{equation}
and
\begin{equation}
(Z_{\|}\Pi_\mu^{\|} \gamma^\mu_{\|}+Z_{\bot}\Pi_\mu^{\bot}
\gamma^\mu_{\bot})E^{l}_{p}(x)=E^{l}_{p}(x)(Z_{\|}\overline{p}^\mu_{\|}
\gamma_\mu^{\|}+Z_{\bot}\overline{p}^\mu_{\bot} \gamma_\mu^{\bot}) \
, \label{eigenproblem-2}
\end{equation}
with $\overline{p}^\mu=(p^{0},0, -\sqrt{2eH l},p^{3})$, thus
$\overline{p}^\mu_{\bot}=(0,0,-\sqrt{2eH l},0)$ and
$\overline{p}^\mu_{\|}=(p^{0},0,0,p^{3})$.

The relations (\ref{eigenproblem})-(\ref{eigenproblem-2}) and the
orthogonality condition (\ref{orthogonality}), facilitate the diagonalization of the fermion self energy
$\Sigma (x,x')$ in momentum space
\begin{eqnarray}\label{P-Self-Energy}
\Sigma(p,p')= \int d^4xd^4y
\overline{E}_{p}^{l}(x)\Sigma(x,y)E_{p'}^{l'}(y)=(2\pi)^4\widehat{\delta}^{(4)}(p-p')\Pi(l)\widetilde{\Sigma}^l
(\overline{p})
\end{eqnarray}
with
\begin{equation}\label{SE-LLL}
\widetilde{\Sigma}^{l}(\overline{p})
=Z_{\|}^{l}\overline{p}_{\|}^\mu\gamma_{\mu}^{\|}+Z_{\bot}^{l}\overline{p}_{\bot}^\mu\gamma_{\mu}^{\bot}+M^{l}I+iT^{l}\gamma^{1}\gamma^{2}
\end{equation}

Using the spin projectors (\ref{Spin-projectors}) and introducing the longitudinal and transverse projectors
\begin{equation}  \label{projector}
\Lambda^{\pm}_{\|}=\frac{1}{2}(1\pm\frac{\gamma^{\|}\cdot
\overline{p}_{\|}}{|\overline{p}_{\|}|}), \qquad
\Lambda^{\pm}_{\bot}=\frac{1}{2}(1\pm i\gamma^2).
\end{equation}
the function $\widetilde{\Sigma}^l (\overline{p})$ can be rewritten in
the following form,
\begin{eqnarray}\label{Sigma-2}
\widetilde{\Sigma}^l
(\overline{p})=Z_{\|}^{l}(\Lambda^{+}_{\|}-\Lambda^{-}_{\|})|\overline{p}_{\|}|
+iZ_{\bot}^{l}(\Lambda^{-}_{\bot}-\Lambda^{+}_{\bot})|\overline{p}_{\bot}|
+(M^{l}+T^{l})\Delta(+)+(M^{l}-T^{l})\Delta(-)
\end{eqnarray}

It is clear from (\ref{P-Self-Energy}) that when the $E_p$
transformation is correctly used (i.e. taking into account the
projector $\Pi(l)$ in the orthogonal condition), the separation between the LLL and the rest of the levels is automatically produced. Considering $l=0$ in (\ref{Sigma-2}), and using that $\overline{p}_{\bot}(l=0)=0$ and $\Delta(+)\Delta(-)=0$, the spinorial structure in the RHS of Eq. (\ref{P-Self-Energy}) reduces to
\begin{equation}\label{Pi-Sigma-zero}
\Pi(0)\widetilde{\Sigma}^0
(\overline{p})=[Z_{\|}^{0}(\Lambda^{+}_{\|}-\Lambda^{-}_{\|})|\overline{p}_{\|}|
+(M^{0}+T^{0})\Delta(+)]\Delta(+)
\end{equation}
While at $l\neq 0$, it is given by
\begin{equation}\label{Pi-Sigma-l}
\Pi(l\neq 0)\widetilde{\Sigma}^l
(\overline{p})=\widetilde{\Sigma}^{l}
(\overline{p})=Z_{\|}^{l}(\Lambda^{+}_{\|}-\Lambda^{-}_{\|})|\overline{p}_{\|}|
+iZ_{\bot}^{l}(\Lambda^{-}_{\bot}-\Lambda^{+}_{\bot})|\overline{p}_{\bot}|
+(M^{l}+T^{l})\Delta(+)+(M^{l}-T^{l})\Delta(-)
\end{equation}

From a physical point of view the results (\ref{Pi-Sigma-zero}) and (\ref{Pi-Sigma-l}) are simply reflecting the fact that there is only one spin projection in the LLL , and hence the self-energy $\widetilde{\Sigma}^0
(\overline{p})$ only contains the spin projector $\Delta(+)$. Since the remaining
LL's contain two spin projections,  $\widetilde{\Sigma}^l
(\overline{p})$ depends on the two projectors $\Delta(+)$ and $\Delta(-)$.

With the help of Eqs.
(\ref{orthogonality}), (\ref{eigenproblem})-(\ref{P-Self-Energy}), it is straightforward to show that the inverse of the full fermion propagator in momentum space is given by
\begin{eqnarray}\label{Full-Inv-Propagator}
G^{-1}_{l}(p,p')= \int d^4xd^4y
\overline{E}_{p}^{l}(x)[\Pi\cdot\gamma -\Sigma(x,y)]E_{p'}^{l'}(y)
=(2\pi)^4\widehat{\delta}^{(4)}(p-p')\Pi(l)[\overline{p}\cdot\gamma
-\widetilde{\Sigma}^l(\overline{p})]
\end{eqnarray}
The full propagator $G^{l}(p,p')$ must satisfy
\begin{eqnarray}\label{Propagator-Eq}
\sum\hspace{-0.47cm}\int \frac{d^{4}p"}{\left( 2\pi \right)
^{4}}G^{-1}_{l"}(p,p")G^{l"}(p",p')=
(2\pi)^4\widehat{\delta}^{(4)}(p-p')\Pi(l)
\end{eqnarray}
where $\sum_{\it l}\hspace{-0.47cm}\int \frac{d^{4}p}{\left( 2\pi
\right) ^{4}}=\sum_{\it l}\frac{dp_0dp_2dp_3}{(2\pi)^4}$. It is easy to see that
(\ref{Propagator-Eq}) is indeed satisfied by
\begin{equation}\label{full-FP}
G^{l}(p,p')=(2\pi)^4\widehat{\delta}^{(4)}(p-p')\Pi(l)\widetilde{G}^{l}(\overline{p})
\end{equation}
with $\widetilde{G}^l(\overline{p})$ formally
given by
\begin{equation}
\widetilde{G}^l(\overline{p})=\frac{1}{\overline{p}\cdot\gamma
-\widetilde{\Sigma}^l(\overline{p})}
\end{equation}\label{Propagator-p}

To find the explicit form of  $\widetilde{G}^l(\overline{p})$ we
have to solve
\begin{equation}\label{inverse}
\widetilde{G}^l(\overline{p})\widetilde{G}^{-1}_l(\overline{p})=\widetilde{G}^{-1}_l(\overline{p})\widetilde{G}^l(\overline{p})=I
\end{equation}
where
\begin{equation}
\widetilde{G}^{-1}_l(\overline{p})=\overline{p}\cdot\gamma
-\widetilde{\Sigma}^l(\overline{p})=\gamma \cdot V^l-M^lI-T^li\gamma_1\gamma_2
\end{equation}\label{Inverse-Prop}
and
$V_\mu=((1-Z_{\|}^{l})p_0,0,(1-Z_{\bot}^{l})\sqrt{2eHl},(1-Z_{\|}^{l})p_3)$.

One can show that the matrix function
\begin{equation}\label{green-func}
\widetilde{G}^{l}(\overline{p})=\frac{AB}{\det
\widetilde{G}^{-1}_l(\overline{p})},
\end{equation}
with
\begin{equation}\label{green-2}
A=\gamma^1\widetilde{G}^{-1}_l(\overline{p})\gamma_1,\quad
B=\gamma_5 \widetilde{G}^{-1}_l(\overline{p})A\gamma_5
\end{equation}
and
\begin{eqnarray}\label{coefficients}
\det \widetilde{G}^{-1}_l(\overline{p})&=&\sqrt[4]{\det
[\widetilde{G}^{-1}_l(\overline{p})AB]}\nonumber
\\
&=&\frac{1}{4}\{[M^l+(V_{\|}^l-T^l)+V_{\bot}^l][M^l-(V_{\|}^l-T^l)+V_{\bot}^l]\nonumber
\\
&+&[M^l+(V_{\|}^l-T^l)-V_{\bot}^l][M^l-(V_{\|}^l-T^l)-V_{\bot}^l]\}\nonumber
\\
&\times&\{[M^l+(V_{\|}^l+T^l)+V_{\bot}^l][M^l-(V_{\|}^l+T^l)+V_{\bot}^l]\nonumber
\\
&+&[M^l+(V_{\|}^l+T^l)-V_{\bot}^l][M^l-(V_{\|}^l+T^l)-V_{\bot}^l]\},
\end{eqnarray}
satisfies the condition (\ref{inverse}).

Using (\ref{green-func}) and working in the basis of projectors (\ref{Spin-projectors}) and (\ref{projector}),
the matrix function $\widetilde{G}^{l}$ defining the full fermion propagator (\ref{full-FP}) can be written as
\begin{eqnarray}\label{Pgorrito}
\widetilde{G}^{l}(\overline{p})
=\frac{N^l(T,V_{\|})}{D^l(T)}\Delta(+)\Lambda^{+}_{\|}
+\frac{N^l(T,-V_{\|})}{D^l(-T)}\Delta(+)\Lambda^{-}_{\|}\qquad\qquad\qquad
\nonumber
\\
+\frac{N^l(-T,V_{\|})}{D^l(-T)}\Delta(-)\Lambda^{+}_{\|}
+\frac{N^l(-T,-V_{\|})}{D^l(T)}\Delta(-)\Lambda^{-}_{\|}\qquad\qquad\qquad\nonumber
\\
-iV_{\bot}^l(\Lambda^{+}_{\bot}-\Lambda^{-}_{\bot})
[\frac{\Delta(+)\Lambda^{+}_{\|}+\Delta(-)\Lambda^{-}_{\|}}{D^l(T)}
+
\frac{\Delta(+)\Lambda^{-}_{\|}+\Delta(-)\Lambda^{+}_{\|}}{D^l(-T)}]\qquad\qquad
\end{eqnarray}
with notation
\begin{eqnarray}\label{coefficients}
N^l(T,V_{\|})&=&T^l-M^l-V_{\|}^l \qquad \qquad \nonumber
\\
D^l(T)&=&(M^l)^2-(V_{\|}^l-T^l)^2+(V_{\bot}^l)^2 \qquad \nonumber
\\
V_{\|}^l&=&(1-Z_{\|}^{l})|\overline{p}_{\|}|\qquad \nonumber
\\
V_{\bot}^l&=&(1-Z_{\bot}^{l})|\overline{p}_{\bot}|=(1-Z_{\bot}^{l})\sqrt{2eHl}.\qquad
\end{eqnarray}

In the LLL,  $V_{\bot}^0=0$, thus the LLL full propagator becomes
\begin{eqnarray}\label{Pgorrito-LLL}
\widetilde{G}^{0}(\overline{p})
=\frac{1}{V^0_{\|}-(M^0+T^0)}\Delta(+)\Lambda^{+}_{\|}
-\frac{1}{V^0_{\|}+(M^0+T^0)}\Delta(+)\Lambda^{-}_{\|}\qquad\qquad\qquad
\nonumber
\\
+\frac{1}{V^0_{\|}-(M^0-T^0)}\Delta(-)\Lambda^{+}_{\|}
-\frac{1}{V^0_{\|}+(M^0-T^0)}\Delta(-)\Lambda^{-}_{\|}\qquad\qquad\qquad
\end{eqnarray}

\section{Schwinger-Dyson Equation for the Fermion Self-Energy }\label{SD-Eq}

To explore the dynamical generation of a magnetic moment in massless
QED, we can start from the SD  equation for the
fermion self-energy in the presence of a constant magnetic field. We
will work in the quenched-ladder approximation where
\begin{equation}
\label{SD} \Sigma (x,x')=-ie^2 \gamma^{\mu}G(x,x')\gamma^{\nu}D_{\mu
\nu}(x-x').
\end{equation}
Here, $\Sigma (x,x')$ is the fermion self-energy operator
(\ref{Mass Operator}), $D_{\mu \nu}(x-x')$ is the bare photon
propagator, and $G(x,x')$ is the full fermion propagator depending
on the dynamically induced quantities and the magnetic field.

Equation (\ref{SD}) can be transformed to momentum space by using
the $E_p$ functions as
\begin{eqnarray}
\label{SD-P} \int
d^4xd^4x'\overline{E}^l_p(x)\Sigma(x,x')E^{l'}_{p'}(x')=-ie^2\int
d^4xd^4x' \overline{E}^l_p(x) \gamma^{\mu} \nonumber
\\
(\sum\hspace{-0.47cm}\int \frac{d^{4}p"}{\left( 2\pi \right)
^{4}}E^{l"}_{p"}(x)\Pi(l")\widetilde{G}^{l^{"}}(\overline{p}")\overline{E}^{l^{"}}_{p"}(x'))\gamma^{\nu}E^{l'}_{p'}(x')
D_{\mu \nu}(x-x')
\end{eqnarray}
where
\begin{equation}  \label{D-Propagator}
D_{\mu
\nu}(x-x')=\int\frac{d^4q}{(2\pi)^4}\frac{e^{-iq\cdot(x-x')}}{q^2-i\epsilon}(g_{\mu
\nu} -(1-\xi )\frac{q_\mu q_\nu}{q^2}),
\end{equation}
with $\xi$ the gauge fixing parameter, and we used that
\begin{eqnarray}\label{Propagator}
G(x,x')= \sum\hspace{-0.49cm}\int \frac{d^{4}p"}{\left( 2\pi
\right) ^{4}}
E_{p"}^{l"}(x)\Pi(l")\widetilde{G}^{l^{"}}(\overline{p}")\overline{E}_{p"}^{l"}(x')
\end{eqnarray}

At this point it is convenient to consider the integrals \cite{Ng}
\begin{eqnarray}\label{vertex-int-1}
\int d^4x \overline{E}^l_p(x) \gamma^{\mu} E^{l"}_{p"}(x)e^{-iq\cdot
x}=(2\pi)^4
\delta^{(3)}(p"+q-p)e^{-iq_1(p_2"+p_2)/2eH}e^{-\widehat{q}^2_\bot/2}\nonumber
\\
\times
\sum\limits_{\sigma,\sigma"}\frac{1}{\sqrt{n!n"!}}e^{i(n-n")\varphi}J_{nn"}(\widehat{q}_\bot
)\Delta(\sigma)\gamma^{\mu}\Delta(\sigma"),\qquad\qquad\qquad
\end{eqnarray}
and
\begin{eqnarray}\label{vertex-int-2}
\int d^4x' \overline{E}^{l"}_{p"}(x') \gamma^{\nu}
E^{l'}_{p'}(x')e^{iq\cdot x'}=(2\pi)^4
\delta^{(3)}(p"+q-p')e^{iq_1(p"_2+p'_2)/2eH}e^{-\widehat{q}^2_\bot/2}\nonumber
\\
\times\sum\limits_{\sigma',\overline{\sigma}"}\frac{1}{\sqrt{n'!\overline{n}"!}}e^{i(\overline{n}"-n')\varphi}J_{\overline{n}"n'}(\widehat{q}_\bot
)\Delta(\overline{\sigma}")\gamma^{\nu}\Delta(\sigma'),\qquad\qquad\qquad
\end{eqnarray}
with $n \equiv n(l,\sigma)$, $n"\equiv n(l",\sigma ")$, $n'\equiv n(l',\sigma')$,
and $\overline{n}" \equiv n(l",\overline{\sigma}")$, defined according to
\begin{equation}  \label{n}
n(l,\sigma)=l+\frac{\sigma}{2}-\frac{1}{2},\qquad l=0,1,2,... ,
\qquad \sigma=\pm 1.
\end{equation}
The notation in (\ref{vertex-int-1}) and (\ref{vertex-int-2}) included the use of polar coordinates for the transverse $q$-momentum $\widehat{q}_\bot\equiv\sqrt{\widehat{q}^{2}_1+ \widehat{q}^{2}_1}$, $\varphi \equiv \arctan (\widehat{q}_2/\widehat{q}_1)$; normalized quantities
$\widehat{Q}_{\mu}=Q_{\mu}/\sqrt{2eH}$; the tri-delta function
\begin{equation}\label{delta3}
\delta^{(3)}(p"+q-p) \equiv \delta(p_0"+q_0-p_0)\delta (p_2"+q_2-p_2)\delta (p_3"+q_3-p_3);
\end{equation}
and
\begin{equation}  \label{n}
J_{nn"}(\widehat{q}_\bot )\equiv
\sum\limits_{m=0}^{min(n.n")}\frac{n!n"!}{m!(n-m)!(n"-m)!}[i\widehat{q}_\bot]^{n+n"-2m}.
\end{equation}
Doing the integrals in $x$ and $x'$ in (\ref{SD-P}) with the help of
(\ref{vertex-int-1}) and (\ref{vertex-int-2}),
integrating in $p"$, and using the Feynman gauge ($\xi =1$), one
finds
\begin{eqnarray}\label{SD-3}
\widetilde{\Sigma}^l(\overline{p}) \Pi(l)\delta^{ll'}=
-ie^2(2eH)\int\frac{d^4\widehat{q}}{(2\pi)^4} \sum\limits_{l"} \sum
\limits_{[\sigma]} \frac{e^{i (n-n"+\overline{n}"-n')\varphi}
}{\sqrt{n!n'!n"!\overline{n}"!}}\frac{e^{-\widehat{q}^2_\bot}}{\widehat{q}^2}\nonumber
\\
\times J_{nn"}(\widehat{q}_\bot
)J_{\overline{n}"n'}(\widehat{q}_\bot
)\Delta(\sigma)\gamma_{\mu}\Delta(\sigma")\Pi(l")\widetilde{G}^{l"}(\overline{p-q})\Delta(\overline{\sigma}")\gamma^{\mu}\Delta(\sigma'),
\end{eqnarray}
where $\overline{p-q} \equiv(p^{0}-q^{0},0,
-\sqrt{2eHl"},p^{3}-q^{3})$ and $[\sigma]$ means
summing over $\sigma, \sigma', \sigma", \overline{\sigma}"$. The appearance of the $\Pi(l)$ factors in both sides of the equation ensures the correct counting of only one spin projection for the fermions at the LLL.

Due to the negative exponential $e^{-\widehat{q}^2_\bot}$ the main
contribution to (\ref{SD-3}) will come from the smallest values of
$\widehat{q}_{\bot}$. This allows one to keep in (\ref{SD-3}) only
the terms with the smallest power of $\widehat{q}_\bot$ in
$J_{\overline{n}"n'}(\widehat{q}_\bot )$ (see Ref. \cite{Ng} for
details). Hence
\begin{equation}  \label{limit}
J_{nn"}(\widehat{q}_\bot )\rightarrow \frac{[max(n',n")]!}{|n-n"|}
[i\widehat{q}_\bot]^{|n-n"|}\rightarrow n!\delta_{nn"}
\end{equation}
and we obtain
\begin{eqnarray}\label{SD-4}
\widetilde{\Sigma}^l(\overline{p}) \Pi(l)\delta^{ll'}=
-ie^2(2eH)\int\frac{d^4\widehat{q}}{(2\pi)^4} \sum\limits_{l"} \sum
\limits_{[\sigma]}
\frac{e^{-\widehat{q}^2_\bot}}{\widehat{q}^2}\qquad\nonumber
\\
\times
\delta_{nn"}\delta_{\overline{n}"n'}\Delta(\sigma)\gamma_{\mu}\Delta(\sigma")\Pi(l")\widetilde{G}^{l"}(\overline{p-q})\Delta(\overline{\sigma}")\gamma^{\mu}\Delta(\sigma'),
\end{eqnarray}

Taking into account that
\begin{equation}  \label{delta}
\delta_{n,n"}=\delta_{l,l"}\delta_{\sigma,\sigma"}+\delta_{l+\sigma,l"}\delta_{-\sigma,\sigma"}
\end{equation}
together with the relations
\begin{equation}  \label{relations-1}
\Delta(\pm)\gamma_{\mu}^{\bot}=\gamma_{\mu}^{\bot}\Delta(\mp),\qquad
\Delta(\pm)\gamma_{\mu}^{\|}=\gamma_{\mu}^{\|}\Delta(\pm)
\end{equation}
\begin{equation}  \label{relations-2}
\Delta(\pm)\Delta(\pm)=\Delta(\pm),\qquad
\Delta(\pm)\Delta(\mp)=0,\qquad
\gamma_{\mu}^{\bot}\gamma_{\nu}^{\bot}\gamma^{\mu}_{\bot}=0,
\end{equation}
we can do the sums in $[\sigma]$ and $l"$ in (\ref{SD-4}) to arrive at the SD
equation
\begin{eqnarray}\label{SD-Eq}
\widetilde{\Sigma}^{l}(\overline{p})\Pi(l)=
-ie^2(2eH)\Pi(l)\int\frac{d^4\widehat{q}}{(2\pi)^4}
\frac{e^{-\widehat{q}^2_\bot}}{\widehat{q}^2}[\gamma_{\mu}^{\|}\widetilde{G}^{l}(\overline{p-q})\gamma_{\mu}^{\|}
\nonumber
\\
+\Delta(+)\gamma_{\mu}^{\bot}\widetilde{G}^{l+1}(\overline{p-q})\gamma_{\mu}^{\bot}\Delta(+)
+
\Delta(-)\gamma_{\mu}^{\bot}\widetilde{G}^{l-1}(\overline{p-q})\gamma_{\mu}^{\bot}\Delta(-)]
\end{eqnarray}

If the external fermion is in the LLL ($l=0$), Eq. (\ref{SD-Eq}) reduces to
\begin{eqnarray}\label{SD-LLL}
\widetilde{\Sigma}^{0}(\overline{p}) \Delta(+)=
-ie^2(2eH)\Delta(+)\int\frac{d^4\widehat{q}}{(2\pi)^4}
\frac{e^{-\widehat{q}^2_\bot}}{\widehat{q}^2}
[\gamma_{\mu}^{\|}\widetilde{G}^{0}(\overline{p-q})\gamma_{\mu}^{\|}
+\gamma_{\mu}^{\bot}\widetilde{G}^{1}(\overline{p-q})\gamma_{\mu}^{\bot}\Delta(+)]\qquad
\end{eqnarray}
with $\widetilde{\Sigma}^{0}(\overline{p}) \Delta(+)$ given in Eq. (\ref{Pi-Sigma-zero}).
While if the external fermion is in any higher LL ($l\neq 0$),  Eq. (\ref{SD-Eq}) becomes
\begin{eqnarray}\label{SD-1LL}
\widetilde{\Sigma}^{(l\neq 0)}(\overline{p})=
-ie^2(2eH)\int\frac{d^4\widehat{q}}{(2\pi)^4}
\frac{e^{-\widehat{q}^2_\bot}}{\widehat{q}^2}[\gamma_{\mu}^{\|}\widetilde{G}^{l}(\overline{p-q})\gamma_{\mu}^{\|}
+\Delta(+)\gamma_{\mu}^{\bot}\widetilde{G}^{l+1}(\overline{p-q})\gamma_{\mu}^{\bot}\Delta(+)
\nonumber
\\
+\Delta(-)\gamma_{\mu}^{\bot}\widetilde{G}^{l-1}(\overline{p-q})\gamma_{\mu}^{\bot}\Delta(-)]\qquad\qquad
\end{eqnarray}
with $\widetilde{\Sigma}^{(l\neq 0)}(\overline{p})$ given in Eq. (\ref{Pi-Sigma-l}).

Since the equation for a given Landau level $l$ involves dynamical
parameters that depend on $l$, $l-1$ and $l+1$, the SD equations for
all the LL's form a system of infinite coupled equations.
Fortunately, in the infrared region, the leading contribution to
each equation will come from the propagators with the lower LL's,
since the magnetic field term $(\sim lB)$ in the denominator of the fermion
propagator for $l\neq 0$ acts as a suppressing factor. Using this
approximation, one can find a consistent solution at each level. It is also convenient to notice that, since the solution for the first LL depends on the one for the LLL; the solution for the second LL depends on the one for the first LL, and so on, the solutions for $M^l$ and $T^l$ can all ultimately be expressed as a function of the LLL solution. This indicates that
the physical origin of all the dynamical quantities is actually due
to the infrared dynamics taking place at the LLL.

\section{Induced Electron mass and Anomalous Magnetic Moment
}\label{induced-quant}

\subsection{Solution of the SD Equation in the LLL}

Let's work in the LLL, that means, $\widetilde{p}_\bot =0$, and consequently the LHS of (\ref{SD-LLL}) is given by (\ref{Pi-Sigma-zero}).
In the infrared region, the leading contribution to the RHS of Eq.
(\ref{SD-LLL}) comes from the term with no magnetic field in the
denominator, that is, the term with the LLL propagator
$\widetilde{G}^{0}(\overline{p-q})$. Hence, in the leading
approximation Eq. (\ref{SD-LLL}) is given by
\begin{eqnarray}\label{SD-LLL-Solution}
(M^0+T^0)\Delta(+)+Z_{\|}^0\Delta(+)(\Lambda^{+}_{\|}-\Lambda^{-}_{\|})|\overline{p}_{\|}|\simeq
-ie^2(2eH)\Delta(+)\int\frac{d^4\widehat{q}}{(2\pi)^4}
\frac{e^{-\widehat{q}^2_\bot}}{\widehat{q}^2}\gamma_{\mu}^{\|}\widetilde{G}^{0}(\overline{p-q})\gamma_{\mu}^{\|}
\end{eqnarray}

Using
(\ref{relations-1})-(\ref{relations-2}) together with
\begin{equation}\label{relations-3}
\gamma_{\mu}^{\|}\Lambda^{\pm}_{\|}\gamma^{\mu}_{\|} =1,
\end{equation}
assuming $Z_{\|}^{(0)}\ll 1$, and doing a Wick's rotation to Euclidean space, we obtain
\begin{eqnarray}\label{P-SD-2}
(M^0+T^0)\Delta(+)+Z_{\|}^0\Delta(+)(\Lambda^{+}_{\|}-\Lambda^{-}_{\|})|\overline{p}_{\|}|=e^{2}(2eH)
\nonumber
\\
\times \Delta(+) \int
\frac{d^4q}{(2\pi)^4}\frac{e^{-\widehat{q}^{2}_{\bot}}}{\widehat{q}^2}
\frac{(M^0+T^0)}{(\overline{p}_{\|}-\overline{q}_{\|})^{2}+(M^0+T^0)^2}
\qquad
\end{eqnarray}

From Eq.(\ref{P-SD-2}) it is clear that $Z_{\|}^0 =0$, which
corroborates the assumption we did before. Taking the infrared limit
$(p_{\parallel}\sim 0)$, and assuming that $M^0+T^0$ is independent
of the parallel momentum, we arrive at
\begin{equation}\label{relations-3}
1=e^2(4eH)\int\frac{d^4\widehat{q}}{(2\pi)^4}
\frac{e^{-\widehat{q}^2_\bot}}{\widehat{q}^2}\frac{1}{(M^{(0)}+T^{(0)})^2+
q_{\|}^2}
\end{equation}
If $M^{0}+T^{0}$ is replaced by $m_{dyn}$, Eq. (\ref{relations-3})
becomes identical to the SD equation obtained in the phenomenon of
magnetic catalysis of chiral symmetry breaking. Thus, the solution of (\ref{relations-3}) is
given by
\begin{equation}
\label{Mass-Eq-Solution} M^{0}+T^{0}\simeq \sqrt{2eH}
e^{-\sqrt{\frac{\pi}{\alpha}}}
\end{equation}
As in \cite{MC-QED}-\cite{Ng}, this solution is obtained considering
that $M^{0}+T^{0}$  does not depend on the parallel momentum, an assumption
consistent within the ladder approximation \cite{BCMA}. As proved in
\cite{Improved-LA}, when the  polarization effect is included in
the gap equation through the improved-ladder approximation, the
solution for $m_{dyn}$ is of the same form as
(\ref{Mass-Eq-Solution}), but with the replacement
$\sqrt{\pi/\alpha} \rightarrow \pi/\alpha \log (\pi/ \alpha)$ in the
exponent. Since the inclusion of the magnetic moment in the LLL SD
equation merely implies the replacement $m_{dyn}\rightarrow
M^{0}+T^{0}$, it is clear that a similar effect will occur in the
solution (\ref{Mass-Eq-Solution}). However, this effect will not
qualitatively change the nature of our findings.

As the fermions in the LLL have only one spin orientation it is not
possible to find $M^{0}$ and $T^{0}$ independently (see that the
combination $M^{0}-T^{0}$ is absent from the LHS of the SD equation (\ref{SD-LLL-Solution}),
as well as from the RHS, because the spin projector $\Delta (+)$ ensures that only the terms containing $M^{0}+T^{0}$ in (\ref{Pgorrito-LLL}) contribute to the RHS of (\ref{SD-LLL-Solution})). Thus, the SD equation determines
the induced LLL rest-energy
\begin{equation}
\label{rest-Energy} E^{0}=M^{0}+T^{0}.
\end{equation}
which has contributions from the dynamical mass and from the
magnetic energy related to the interaction between the magnetic field and the dynamically induced magnetic moment.

\subsection{Solution of the SD Equation in the First-LL}

The SD equation for a fermion in the first-LL is
\begin{eqnarray}\label{SD-N-LLL}
\widetilde{\Sigma}^{1}(\overline{p})=
-ie^2(2eH)\int\frac{d^4\widehat{q}}{(2\pi)^4}
\frac{e^{-\widehat{q}^2_\bot}}{\widehat{q}^2}[\gamma_{\mu}^{\|}\widetilde{G}^{1}(\overline{p-q})\gamma_{\mu}^{\|}+\qquad\nonumber
\\
+\Delta(+)\gamma_{\mu}^{\bot}\widetilde{G}^{2}(\overline{p-q})\gamma_{\mu}^{\bot}\Delta(+)
+\Delta(-)\gamma_{\mu}^{\bot}\widetilde{G}^{0}(\overline{p-q})\gamma_{\mu}^{\bot}\Delta(-)]\qquad
\end{eqnarray}

Here again the leading contribution comes from the term with no
magnetic field in the denominator. Hence to find the leading
contribution we just need to keep the term depending on
$\widetilde{G}^{0}(\overline{p-q})$ in the RHS. Using
(\ref{relations-1})-(\ref{relations-2}), together with
\begin{equation}\label{relations-4}
\gamma_{\mu}^{\bot}\Lambda^{\pm}_{\|}\gamma^{\mu}_{\bot}
=2\Lambda^{\mp}_{\|},
\end{equation}
and working as before in the infrared limit, the SD equation after Wick's rotation
becomes
\begin{eqnarray}\label{SD-1-LL}
Z_{\bot}^{1}\gamma_2(2eH)+(M^{1}+T^{1})\Delta(+)+(M^{1}-T^{1})\Delta(-)
= e^2(4eH)\Delta(-)\int\frac{d^4\widehat{q}}{(2\pi)^4}
\frac{e^{-\widehat{q}^2_\bot}}{\widehat{q}^2}\frac{E^{0}}{(E^{0})^2+q_{\|}^2}\qquad\qquad
\end{eqnarray}
Therefore,
\begin{equation}\label{dif-sum}
M^{1}+T^{1}=0,\quad Z_{\bot}^{1}=0
\end{equation}
and
\begin{equation}\label{1-LL-Eq}
M^{1}-T^{1}=e^2(4eH)\int\frac{d^4\widehat{q}}{(2\pi)^4}
\frac{e^{-\widehat{q}^2_\bot}}{\widehat{q}^2}\frac{E^{0}}{(E^{0})^2-q_{\|}^2}
\end{equation}
Taking into account (\ref{relations-3}) in (\ref{1-LL-Eq}) we get
\begin{eqnarray}\label{dif-sum-2}
M^{1}-T^{1}=E^{0}
\end{eqnarray}

Finally, from (\ref{dif-sum}) and (\ref{dif-sum-2}), it results
\begin{equation}\label{M}
M^{1}=-T^{1}=\frac{1}{2} E^{0}=\sqrt{eH/2}
e^{-\sqrt{\frac{\pi}{\alpha}}},\qquad
\end{equation}

The solution (\ref{M}) corroborates the relevance of the LLL dynamics (both
$M^{1}$ and $T^{1}$ are determined by $E^{0}$) in the generation of
the dynamical mass and magnetic moment of the fermions in the first
LL. Given that the magnitude of the magnetic moment for the
electrons in the first LL is determined by the dynamically generated
rest-energy of the electrons in the LLL, any modification of the
theory producing an increase of $E^{0}$ will, in turn, lead to an
increase in the magnitude of $T^{1}$. From the experience with the
$MC\chi SB$ phenomenon, such modifications could be for example,
lowering the space dimension \cite{Gusynin}, introducing
scalar-fermion interactions \cite{Vivian, BCMA}, or considering a
non-zero bare mass \cite{Mass}.

For the remaining LL's the procedure is similar. For
example, the leading
term in the second LL will be given by the $\widetilde{G}^1(\overline{p-q})$
contribution which in turn depends on $E_0$ through the values found for $M^1$ and $T^1$.
Therefore, the values of $M^{l}$ and $T^{l}$ for higher LL's
will depend on $E^{0}$ through the found values of the previous LL's. This
fact shows that the infra-red dynamics of the electrons in the
LLL is the dominant one.

\section{Dispersion Relations and Zeeman Splitting}\label{zeeman-split}

\subsection{Fermions in the LLL}
In the Appendix we showed that the Dirac equation for fermions in the LLL can be written as
\begin{equation}
\label{LLL-electron-eq}
[\widetilde{p}\cdot\widetilde{\gamma}-E^0]\psi_{LLL} =0,
\end{equation}
where $\psi_{LLL}$ is a spin-up two-component spinor. Eq.
(\ref{LLL-electron-eq}) coincides with that of the free (1+1)-D
Thirring model \cite{Th-M}, with the $(1+1)-D$ gamma matrices
$\widetilde{\gamma}_{0}=\sigma_1$,
$\widetilde{\gamma}_{1}=-i\sigma_2$, defined in terms of the Pauli matrices $\sigma_i$, and $\widetilde{p}_{\mu}=(p_0,p_3)$.

The dispersion relation of the LLL fermions obtained from
(\ref{LLL-electron-eq}) is
\begin{equation}
\label{LLL-dispers-eq} p_0=\pm \sqrt{p_3^2+(E^0)^2},
\end{equation}
Thus, the effect of a dynamical magnetic moment is
irrelevant for the LLL fermions, since it just redefines their rest energy through the replacement
$m_{dyn}\rightarrow M^{0}+T^{0}$. This is physically natural, since
the fermions in the LLL can only have one spin projection, so for
them there is no spin degeneracy and hence, no possible energy
splitting due to the magnetic moment. We shall see below that the dynamical anomalous magnetic moment turns out to be really relevant for fermions in higher LL's.
\subsection{Fermions in the First-LL}

Let us find now the dispersion relations for fermions in higher
LL's, taking into account the dynamically induced quantities.
Starting from the modified field equation in the presence of the
magnetic field,
\begin{equation}
\label{electron-eq}
[\overline{p}\cdot\gamma-M^lI-iT^l\gamma^1\gamma^2]\psi_l =0,
\end{equation}
where we neglected the coefficients
$Z$ in the terms $(1-Z^l_{\|})$ and $(1-Z^l_{\bot})$, since it is
expected that the $Z's$ are much smaller than one; the dispersion relations are found from
\begin{eqnarray}
\label{determinant}
&det&[\overline{p}\cdot\gamma-M^lI-iT^l\gamma^1\gamma^2]=
[(M^l)^2-(\overline{p}_{\|}-T^l)^2+\overline{p}_{\bot}^2]
[(M^l)^2-(\overline{p}_{\|}+T^l)^2+\overline{p}_{\bot}^2] =0.
\end{eqnarray}
yielding
\begin{equation}
\label{disp-relat-1} p_{0}^{2}=p_{3}^{2}+[\sqrt{(M^l)^{2}+2eHl}\pm
T^l]^2,
\end{equation}
and thus showing that the induced magnetic moment breaks the energy
degeneracy between the spin states in the same LL (see the double sign in front of $T^l$).

In particular for the first-LL, and taking into account
that $M^{1}/\sqrt{2eH}\ , T^{1}/\sqrt{2eH} \ll 1$, the leading contribution to the energy becomes
\begin{equation}\label{disp-relat-3}
p_{0}^{2}\simeq p_{3}^{2}+2eH+(M^1)^2+(T^1)^2\pm2T^{1}\sqrt{2eH},
\end{equation}

Working in the infrared region
($p_3^2/2eH \ll 1$) the expression (\ref{disp-relat-3}) can be approximated as
\begin{equation}\label{energy}
p_{0}\simeq \pm
[\sqrt{2eH}+\frac{p_{3}^2}{2\sqrt{2eH}}+\frac{(M^1)^2+(T^1)^2}{2\sqrt{2eH}}\pm
T^{1}]
\end{equation}

As a consequence, the energy splitting for the fermions in the first LL is
\begin{equation}\label{energy-splitting}
\Delta E=|2T^{1}|=2 \sqrt{eH/2} e^{-\sqrt{\pi /\alpha}}
\end{equation}
One can rewrite the last expression in the usual form of the Zeeman
energy splitting for the two spin projections already introduced in
Sec. \ref{zeeman-eff-gener} in Eq. (\ref{Zeeman-splitting}) as
$\Delta E=\widetilde{g}\widetilde{\mu}_B H$
with $\widetilde{g}$ and $\widetilde{\mu}_B$ representing the
non-perturbative Lande $g$-factor and Bohr magneton respectively given by
$\widetilde{g}=2e^{-2\sqrt{\pi /\alpha}},\quad \widetilde{\mu}_B
=\frac{e}{2M^1}$.

\section{Chiral Condensate}\label{Chiral-Con}
Now we are going to find the relation between the chiral condensate $\langle \overline{\Psi}\Psi \rangle$, that is, the order parameter of the magnetically catalyzed chiral symmetry breaking, and the LLL fermion rest-energy, which as seen before depends on the sum of the LLL mass and magnetic moment energy contribution.

We start from the definition of the chiral condensate
\begin{equation}\label{condensate}
\langle \overline{\Psi}\Psi \rangle=iTr[G(x,x)/V],
\end{equation}
with $G(x,x)$ the full fermion propagator given in (\ref{Propagator}), and $V$ the system volume. As discussed before, the phenomenon of magnetic catalysis of chiral symmetry breaking is physically due to the infrared dynamics of the fermions in the LLL. Thus, the leading contribution to the condensate (\ref{condensate}) comes from the LLL fermions. It is convenient to recall here that the space-dependent part of the LLL fermion wave function is given by \cite{Sokolov}
\begin{equation}\label{magnetic-length}
\Psi(x)\sim e^{i(x^0p_0+x^2p_2+x^3p_3)}e^{-\frac{(x_1-x_c)^2}{4l^2_B}},
\end{equation}
where $x_c=p_2l^2_B$ is the coordinate of the center of the Landau orbits and $l_B=1/\sqrt{eB}$ is the magnetic length. It is clear from (\ref{magnetic-length}) that a particle in the LLL can be localized along the $x_2$ and $x_3$ directions up to infinite, but along the $x_1$ direction it is confined within a magnetic length due to the Gaussian function with width $l_B$ (note that for (\ref{magnetic-length}) the standard deviation from the particle position average value $\langle x_1 \rangle$ is $\sigma= l_B$). This implies that in the LLL the volume $V$ in (\ref{condensate}) is given by $V=L_0L_2L_3l_B$, with $L_i=\int_{-\infty}^{+\infty} dx_i$.

Using (\ref{Propagator}) in (\ref{condensate}) and keeping only the leading LLL contribution in the sum, we obtain after integrating in $x's$,
\begin{equation}\label{condensate-LLL}
\langle \overline{\Psi}\Psi \rangle=i\frac{L_0L_2L_3}{V}tr\{\int \frac{dp_0dp_2dp_3}{(2\pi)^3}\Delta (+)\widetilde{G}^0(\overline{p})\}
\end{equation}
with $\widetilde{G}^0(\overline{p})$ given in (\ref{Pgorrito-LLL}), and $\textit{tr}$ denoting the remaining spinorial trace. In the above result we used the orthogonality of the parabolic cylinder functions
\begin{equation}\label{Orthogonality-PCF}
\int_{-\infty}^{\infty}d\rho D_l(\rho)D_{l'}(\rho)=\sqrt{2\pi}l!\delta_{ll'}
\end{equation}
The trace operation reduces the previous expression to
\begin{equation}\label{condensate-LLL-2}
\langle \overline{\Psi}\Psi \rangle=i\frac{2}{l_B}\int \frac{dp_0dp_2dp_3}{(2\pi)^3}\frac{E^0}{(E^0)^2+|p_\||^2}
\end{equation}
Transforming to Euclidean space, taking polar coordinates for the parallel momenta, and using a momentum cut-off defined by the magnetic scale $1/l_B$, which is dominant in the infra-red region, we have
\begin{equation}\label{condensate-LLL-2}
\langle \overline{\Psi}\Psi \rangle=-\frac{1}{(2\pi)^2l_B}\int_{-\frac{1}{l_B}}^{\frac{1}{l_B}}dp_2\int_0^{\frac{1}{l_B}} dp_{\|}^2\frac{E^0}{(E^0)^2+|p_\||^2}
\end{equation}
After integrating in the momenta we obtain
\begin{equation}\label{condensate-2}
\langle \overline{\Psi}\Psi \rangle\simeq -\frac{eH}{2\pi^{2}}E^{0}\ln(\frac{eH}{(E^{0})^{2}})
\end{equation}

This result shows that the role played by the dynamical mass in previous works on magnetic catalysis, on which the induced anomalous magnetic moment was ignored \cite{Ng}, is now played by $E^0$. Considering the magnetic-field dependence of $E^0$ given in (\ref{Mass-Eq-Solution}), we can rewrite (\ref{condensate-2}) as
\begin{equation}\label{condensate-3}
E^{0}=\frac{2\pi^{2}l_B^2}{\ln 2-\sqrt{\pi/\alpha}}\langle \overline{\Psi}\Psi \rangle
\end{equation}
It shows that the induced rest-energy of the electrons in the LLL is proportional to the condensate. This reflects the fact that the same order parameter produces the induction of the two quantities contributing to $E^0$: the dynamical mass and anomalous magnetic moment. This also confirms that the dynamical mass and magnetic moment have a common physical origin, as they are related to one single order parameter, the chiral condensate $\langle \overline{\Psi}\Psi \rangle$.

This validates the fact that since  the induction of the term $\sigma_{\mu\nu}\widehat{F}^{\mu\nu}$ in the electron self-energy does not produce any additional symmetry breaking to that already created by the mass term, then once the chiral symmetry is broken by the condensate $\langle \overline{\Psi}\Psi \rangle$, both dynamical parameters $M^0$ and $T^0$ are generated.

\section{Concluding Remarks}\label{con-remarks}
In this paper we have presented a fresh view of the phenomenon of magnetic catalysis of chiral symmetry breaking in massless QED. Our results show that in this phenomenon the dynamical generation of an additional parameter: the fermion anomalous magnetic moment, has to be considered in equal footing to that already taken into account in previous works on this topic: the fermion mass. The rationale for the increase in the number of the induced parameters is easy to understand on symmetry arguments. The reason is that once the chiral symmetry is broken by the magnetically catalyzed condensation of fermion/anti-fermion pairs, all the physical quantities in the action that were forbidden by the chiral symmetry of the bare theory can be now dynamically induced. Therefore, the chiral condensate induces dynamical mass and anomalous magnetic moment, since in the massless theory their generation was only protected by that same symmetry. Another important outcome of this paper is the lifting of the spin degeneracy of all the LLs different from zero. The degeneracy is eliminated due to the Zeeman splitting produced by the dynamical anomalous magnetic moment. This energy splitting is given in terms of a non-perturbative Lande $g$-factor and a Bohr magneton that depends on the dynamical mass.

We call attention to a fact that has not been emphasized in previous works on magnetic catalysis,
but that is really essential to prove the results here obtained. We refer to the dependence of the dynamical quantities on the Landau levels. There is no reason to assume a priori that the dynamical parameters have to be the same for all the LLs. One can see that, as soon as one is interested in exploring the SD equations for higher LLs. It becomes clear that there is no consistent solution of the infinite system of coupled equations, unless one takes into consideration that the mass and the magnetic moment depend on the LLs. This is evident already from Eq. (\ref{SD-1-LL}).

We anticipate that the dynamical generation of the magnetic moment will be a universal feature of theories with magnetic catalysis, in the same way that occurs with the generation of the dynamical mass. Moreover, we expect that an anomalous magnetic moment will be also dynamically generated in theories with fermion/fermion condensates in the presence of a magnetic field, as long as the symmetry broken by the condensate coincides with the one that would be explicitly broken by a magnetic moment term in the action. These considerations point to some potential applications of our findings in two different areas: condensed matter and the ultra dense matter existing in the core of compact stars.

In condensed matter, the most plausible application at present could be in the physics of graphene. It is known that
the 2-dimensional crystalline form of carbon known as graphene \cite{graphene} has charge carriers that behave as
massless Dirac electrons. In particular, a phenomenon where the
dynamically induced Zeeman effect we have found here could shed some new light is the
lifting of the fourfold degeneracy of the $l=0$ LL, and twofold
degeneracy of the $l=1$ LL in the recently found quantum Hall states
corresponding to filling factors $\nu =0,\pm 1, \pm 4$ under strong
magnetic fields \cite{QH}. An attempt to explain the observed lifting of the LLs degeneracy was carried out in \cite{Gorbar} with the help of a 2+1-dimensional four-fermion-interaction model of Dirac quasiparticles with chemical potentials interpreted as Quantum Hall Ferromagnetism (QFH) order parameters, and dynamical masses related to the phenomenon of $MC$$\chi$$SB$. Nevertheless, we should call attention to two shortcomings of Ref. \cite{Gorbar}. First,  the order parameter $\widetilde{\mu}_s$ was introduced in \cite{Gorbar} as a chemical potential and interpreted as a QFH order parameter, implying that its physical origin was considered unrelated to the phenomenon of magnetic catalysis. However, it is not hard to understand, based on the results of the present paper, that  $\widetilde{\mu}_s$ should have been identified as a 2+1 dynamical anomalous magnetic moment,  thus similar to the one we have found in 3+1 massless QED. In other words,  $\widetilde{\mu}_s$ should have been connected to the $MC$$\chi$$SB$ phenomenon.  Second, the LLL results reported in the second paper of \cite{Gorbar} are not self-consistent, since they were found taking into account the single (pseudo)spin contribution in the LLL on the RHS of the gap equation (A31), but ignoring it in the self-energy operator appearing in the LHS of the same equation. As a consequence, separated values for the LLL parameters analogous to our $M^0 $ and $T^0$  were obtained.

The other possible application in the realm of ultra-dense matter relies on the phenomenon of color superconductivity. An important aspect of color
superconductivity is its magnetic properties
\cite{alf-raj-wil-99/537}-\cite{Vortex}. In spin-zero color
superconductivity, although the color condensate has non-zero
electric charge, it is neutral with respect to a modified electromagnetism. This so-called rotated electromagnetism is present because a linear combination of the photon and the eight gluon remains massless, hence giving rise, in both the 2SC and
CFL phases, to a long-range "rotated-electromagnetic" field
\cite{alf-raj-wil-99/537}. The long-range field can propagate in the color superconductor implying that there is no Meissner effect for the rotated component of an external magnetic field applied to the color superconductor. Even though the quark-quark condensate is
neutral with respect to the rotated charge, an applied magnetic field can interact with the quarks of a pair formed by $\widetilde{Q}$-charged quarks of opposite sign and, moreover, for large magnetic fields this interaction can reinforce these pairs \cite{MCFL}. In this sense, the "rotated-electromagnetism" in the
color-superconductor has some resemblance with the chiral condensate in a theory with magnetic catalysis. It is natural to expect then that a dynamically magnetic moment can also be induced in a color
superconductor under an applied magnetic field. Since, on the other
hand, the Meissner instabilities that appear in some density regions
of the color superconductor can be removed by the induction of a
magnetic field \cite{Vortex}, it will be interesting to investigate
what could be the role in this process of a dynamically induced
magnetic moment.

\textbf{Acknowledgments}

This work was supported in part by the Office of Nuclear Physics of the Department of Energy under contract DE-FG02-09ER41599.

\appendix

\section{The LLL Lagrangian in the Chiral-Condensate Phase}
\label{DE-LLL}

We are interested in obtaining the Dirac equation in momentum space
for fermions in the LLL of the chiral-condensate phase (\ref{LLL-electron-eq}). With this goal in mind,
we should start from the Dirac Lagrangian in the presence of a constant and uniform magnetic field that includes the self-energy corrections
\begin{equation}\label{A8}
{\cal L}=\int d^4x\overline{\psi}(x)(\Pi_\mu \gamma^\mu -
\Sigma(x))\psi(x)
\end{equation}

Using Ritus' transformation to momentum space for the wave
functions
\begin{eqnarray}\label{A9}
\psi(x)=\sum\hspace{-0.47cm}\int \frac{d^{4}p}{\left( 2\pi
\right) ^{4}}E_p^l(x)\psi_l(p),\qquad \overline{\psi}(x)=\sum \hspace{-0.47cm}\int \frac{d^{4}p'}{\left( 2\pi \right)
^{4}}\overline{\psi}_{l'}(p')\overline{E}_{p'}^{l'}(x)
\end{eqnarray}

and taking into account (\ref{orthogonality}), (\ref{eigenproblem}) and
(\ref{P-Self-Energy}), we obtain
\begin{equation}\label{A10}
{\cal L}=\sum \hspace{-0.47cm}\int \frac{d^{4}p}{\left( 2\pi
\right) ^{4}}\overline{\psi}_l(p)\Pi(l)[ \gamma^\mu\overline{p}_\mu
- \widetilde{\Sigma}^l(\overline{p})]\psi_l(p)
\end{equation}

The factor $\Pi(l)$, given in Eq. (\ref{degeneracy}), separates
the LLL Lagrangian ${\cal L}_{0}$ from the rest

\begin{equation}\label{A11}
{\cal L}={\cal L}_{0}+\sum_{l=1}^{\infty}\int \frac{dp_0dp_2dp_3}{\left(
2\pi \right) ^{4}}\overline{\psi}_l(p)[ \gamma^\mu\overline{p}_\mu -
\widetilde{\Sigma}^l(\overline{p})]\psi_l(p)
\end{equation}
with
\begin{equation}\label{A10-1}
{\cal L}_{0}=\int \frac{dp_0dp_2dp_3}{\left( 2\pi \right)
^{4}}\overline{\psi}_0(p)\Delta(+)[ \gamma^{\|}\cdot p_{\|} -
\widetilde{\Sigma}^0(\overline{p})]\psi_0(p)
\end{equation}

Considering the Dirac matrices in the chiral representation
\begin{equation}\label{A1}
\gamma^0=\beta =  \,\left(
\begin{array}{cc}
 0
  & -1 \\
-1 & 0
\end{array}\right) \ ,\qquad
\gamma^i=  \,\left(
\begin{array}{cc}
 0
  & \sigma^i \\
-\sigma^i & 0
\end{array}\right) \ ,\qquad
\gamma^5=  \,\left(
\begin{array}{cc}
 1
  &  \\
 & -1
\end{array}\right) \ ,
\end{equation}
where $\sigma^i$ are the Pauli matrices, we can express the spin
projectors (\ref{Spin-projectors}) in terms of $\sigma^i$ as
\begin{equation}\label{A2}
\Delta(\pm) =  \,\left(
\begin{array}{cc}
 \sigma^{\pm}
  &  \\
 & \sigma^{\pm}
\end{array}\right) \ ,
\end{equation}
with
\begin{equation}\label{A3}
\sigma^{\pm}=\frac{1}{2}(1\pm\sigma^3),
\end{equation}
and introduce the chiral projection operators
\begin{equation}\label{A4}
R= \frac{1+ \gamma _{5}}{2}= \,\left(
\begin{array}{cc}
 1
  &  \\
 & 0
\end{array}\right), \,\qquad
  L= \frac{1- \gamma _{5}}{2}= \,\left(
\begin{array}{cc}
 0
  &  \\
 & 1
\end{array}\right) \ .
\end{equation}
 The projectors (\ref{A2}) and (\ref{A4}) satisfy the commutation relations
\begin{equation}\label{A5}
[\Delta(\pm),L]=[\Delta(\pm),R]=0
\end{equation}
Introducing now the chiral-spin representation for the Dirac spinor
\begin{equation}\label{A5}
\psi_R^{(+)}=R\Delta(+)\psi,\quad \psi_R^{(-)}=R\Delta(-)\psi,\quad
\psi_L^{(+)}=L\Delta(+)\psi,\quad \psi_L^{(-)}=L\Delta(-)\psi
\end{equation}
\begin{equation}\label{A6}
\overline{\psi}_R^{(+)}=\overline{\psi}L\Delta(+),\quad
\overline{\psi}_R^{(-)}=\overline{\psi}L\Delta(-),\quad
\overline{\psi}_L^{(+)}=\overline{\psi}R\Delta(+),\quad
\overline{\psi}_L^{(-)}=\overline{\psi}R\Delta(-)
\end{equation}
and using that
\begin{equation}\label{A7}
\psi=\psi_R^{(+)}+\psi_R^{(-)}+ \psi_L^{(+)}+\psi_L^{(-)}\qquad
\end{equation}
we obtain
\begin{equation}\label{A20}
{\cal L}_{0}=\int \frac{dp_0dp_2dp_3}{\left( 2\pi \right)
^{4}}[\overline{\psi}_{0R}^{(+)}(p)\gamma^{\|}\cdot p_{\|}
\psi_{0R}^{(+)}(p)-\overline{\psi}_{0R}^{(+)}(p)E^0\psi_{0L}^{(+)}(p)+\overline{\psi}_{0L}^{(+)}(p)\gamma^{\|}\cdot
p_{\|}
\psi_{0L}^{(+)}(p)-\overline{\psi}_{0L}^{(+)}(p)E^0\psi_{0R}^{(+)}(p)]
\end{equation}
where we used $\Delta(+)\widetilde{\Sigma}^0(\overline{p})= E^0\Delta(+)$ since $Z_{\|}^0=0$ (see Eq. (\ref{Pi-Sigma-zero})).
From (\ref{A20}) we see that the LLL only gets contribution from the wave functions of the spin up-states. Introducing $\psi^\top_0=(\psi_1,\psi_2,\psi_3,\psi_4)$ for the LLL four spinors we can rewrite Eq. (\ref{A20}) as
\begin{equation}\label{A22}
{\cal L}_{0} = \int \frac{dp_0dp_2dp_3}{\left( 2\pi \right) ^{4}} \
(\psi_3^\ast,\psi_1^\ast)  \left(
\begin{array}{cc}
 E_0
  & (p_0+p_3) \\
(p_0-p_3) & E_0
\end{array}\right)
\left(
\begin{array}{c}
\psi _1 \\
\psi _3
\end{array}
\right) \ ,
\end{equation}
Notice that in the LLL the fermion spinor reduces to a bispinor which corresponds to the two chiralities of the spin up state in
this case.

At this point it is more convenient to work with the (1+1)-D gamma matrices
\begin{equation}\label{A23}
\widetilde{\gamma}^0= \sigma_1= \,\left(
\begin{array}{cc}
 0
  & 1 \\
 1& 0
\end{array}\right) \ ,
\quad  \widetilde{\gamma}^1= -i\sigma_2= \,\left(
\begin{array}{cc}
 0
  & -1 \\
 1& 0
\end{array}\right) \
\end{equation}
which satisfy the algebra
\begin{equation}\label{A23-1}
\widetilde{\gamma}^\mu \widetilde{\gamma}^\nu=g^{\mu
\nu}+\epsilon^{\mu \nu}\widetilde{\gamma}^\nu,
\end{equation}
\begin{equation}\label{A23-2}
\widetilde{\gamma}^\mu \widetilde{\gamma}^5=-\epsilon^{\mu
\nu}\widetilde{\gamma}_\nu
\end{equation}
with $\widetilde{\gamma}^5=\widetilde{\gamma}^0\widetilde{\gamma}^1$ and the (1+1)-D metric and the totally antisymmetric tensor given respectively by
\begin{equation}\label{A24}
g^{\mu\nu}=\left(
\begin{array}{cc}
1 &  \\
& -1
\end{array}
\right) \ ,\quad \epsilon^{\mu\nu}=\left(
\begin{array}{cc}
0 & 1 \\
-1 & 0
\end{array}
\right)
\end{equation}
Defining the LLL bi-spinor as
\begin{equation}\label{A24}
\psi _{LLL}= \left(
\begin{array}{c}
\psi _1 \\
\psi _3
\end{array}
\right)
\end{equation}
and using (\ref{A23}), Eq. (\ref{A22}) can be rewritten in the compact form
\begin{equation}\label{A25}
{\cal L}_{0}=\int \frac{dp_0dp_2dp_3}{\left( 2\pi \right) ^{4}} \
\overline{\psi}_{LLL}(p) \ [ \widetilde{\gamma} \cdot \widetilde{p}-
E^0] \ \psi_{LLL}(p)
\end{equation}
where $\widetilde{p}_{\mu}=(p_0,p_3)$. From the Lagrangian (\ref{A25}) we obtain the Dirac equation
for the LLL fermions
\begin{equation}
\label{LLL-DE}
[\widetilde{p}\cdot\widetilde{\gamma}-E^0]\psi_{LLL} =0,
\end{equation}

The Hamiltonian associated to ${\cal L}_{0}$ is
\begin{equation}\label{A26}
{\cal H}=E^0+p_3\widetilde{\gamma}^1
\end{equation}
which is the free (1+1)-D Thirring model Hamiltonian \cite{Th-M}
with the replacement of the fermion mass $m$ by the rest-energy
$E^0$. Thus, in the chiral-condensate phase the LLL fermions behave as free particles in a reduced $(1+1)$-D space with rest energy proportional to the inverse magnetic length $\textit{l}_{\textit{mag}}^{-1}=\sqrt{eH}$ (see Eq.  (\ref{Mass-Eq-Solution})).

\end{document}